\documentclass[aps,prl,reprint,superscriptaddress]{revtex4-1}
\usepackage{graphicx}
\usepackage{color}
\usepackage{multirow}

\usepackage{hyperref}
\usepackage[inline]{trackchanges}
\include{ds-revision-tools-202107.tex}
\begin{document}

%\preprint{}

\title{Deep learning for CALPHAD modeling: \\Universal parameter learning solely based on chemical formula}
%\title{Deep Learning for Calphad Modeling: All Data Training}
%\title{Deep Learning for Calphad Modeling: Framework}



\author{Qi-Jun Hong}
\email[]{qijun.hong@asu.edu}
\affiliation{School for Engineering of Matter, Transport, and Energy, Arizona State University, Tempe, Arizona 85285, USA}


\date{\today}

\begin{abstract}
Empowering the creation of thermodynamic and property databases, the CALPHAD (CALculation of PHAse Diagrams) methodology plays a vital role in enhancing materials and manufacturing process design. In this study, we propose a deep learning approach to train parameters in CALPHAD models solely based on chemical formula. We demonstrate its application through an example of calculating the mixing parameter of liquids. This work showcases the integration of CALPHAD and deep learning, highlighting its potential for achieving automated comprehensive CALPHAD modeling.

\pacs{}

\keywords{deep learning, CALPHAD modeling, mixing parameter}

\end{abstract}



\maketitle

%\section{}
%\subsection{}
%\subsubsection{}

\section{ \textbf{Introduction} }

A comprehensive understanding of phase equilibria and the underlying thermodynamics is essential for studying materials and their transformations, and thus phase diagrams serve as valuable roadmaps for materials development. The graphical representation of systems with more than two or three components poses challenges, which restricts their usefulness, and multicomponent systems often have incomplete phase diagram information. Computational thermodynamics methods offer a promising solution to address these limitations. The groundwork for the CALPHAD (CALculation of PHAse Diagrams) method, as we know it today, was laid by Kaufman and Bernstein \cite{Kaufman1970} in 1970. Their seminal book provided an overview of the method's fundamental features, including the publication of computer programs for calculating binary and ternary systems. The temperature dependence of Gibbs free energy $G(T)$ is usually expressed as a series of temperature terms,
\begin{equation}
G(T) = a + bT + cT\ln(T) + \sum_n{d_nT^n},
\end{equation}
where $a$, $b$, $c$ and $d_n$ are coefficients and $n$ are integers. The composition dependence of the Gibbs energy is described as 
\begin{eqnarray}
G(A_{x_A}B_{x_B}) = &&x_A G_A^{\circ} + x_B G_B^{\circ} \nonumber \\
&&+ RT\left( x_A \ln{x_A} + x_B \ln{x_B} \right) \nonumber \\
&&+ x_A x_B \sum_{i=0}^{n_{AB}}{L_{A,B;i}(x_A - x_B)^i}, 
\end{eqnarray}
where $x_A$ and $x_B$ are the mole fractions of elements A and B, $G_A^{\circ}$ and $G_B^{\circ}$ are Gibbs free energies of the elements and $L_{A,B;i}$ are the excess parameters. 

Traditionally, the model parameters of Gibbs energy functions have been derived from experimental data or computational data obtained through atomistic methods. Optimized model parameter sets are obtained by considering phase diagram and thermochemical data to fit different types of data. Mathematical techniques like the least-squares method of Gauss \cite{Lukas1977}, the Levenberg-Marquardt method \cite{Marquardt1963}, or Bayesian estimation method \cite{Konigsberger1991} are employed to adjust the model parameters and efficiently fit critically evaluated data. These methods have been successfully implemented in various CALPHAD software.

Despite the progress, current CALPHAD modeling still relies on a highly ``piecewise" approach, where each parameter fitting process is constrained to a particular system of interest, without harvesting the data outside of the system. In this work, we aim to integrate CALPHAD modeling and deep learning. As a first step, we leverage the graph neural network (GNN) method we developed recently \cite{hong2021melting, Hong2022_pnas, Hong2022_cms, Hong2022_calphad} to propose a universal approach to estimate CALPHAD parameters, with an ultimate goal to learn and determine CALPHAD parameters through deep learning of the entire CALPHAD data available. 

We describe our approach to learn CALPHAD parameters, including GNN, ensemble model, and multi-task learning in the Methods section. We present an example of learning the first-order excess parameter of liquids ($L_{A,B;0}$) in Results section.

\section{Methods \label{methods}}
We utlize our recently developed GNN approach to establish a mapping from chemical formula to CALPHAD parameters. The advantage of this approach is its minimal input requirement, solely relying on the chemical formula. This simplicity enables the extensive future application of the model, as it does not require any additional materials properties, and thus neither computational nor experiment data are needed. In our view, relying on input feature would reduce the broad applicability of the method, especially in obtaining unknown systems, where chemistry is typically the only a priori known input. 

The machine learning model presented in this study combines the GNN \cite{Scarselli2009} and residual neural network (ResNet) \cite{He2016} architectures, implemented within the Tensorflow \cite{tensorflow} framework. We specifically designed the GNN architecture to enforce permutation invariance (e.g., ZrO$_2$ and O$_2$Zr are the same material), thereby reducing model complexity and enhancing efficiency. Additionally, the ResNet architecture addresses the issue of vanishing gradient through the use of skip connections, further streamlining the network's design. 

\begin{figure}[]
    \centering
    \includegraphics[width=0.49\textwidth]{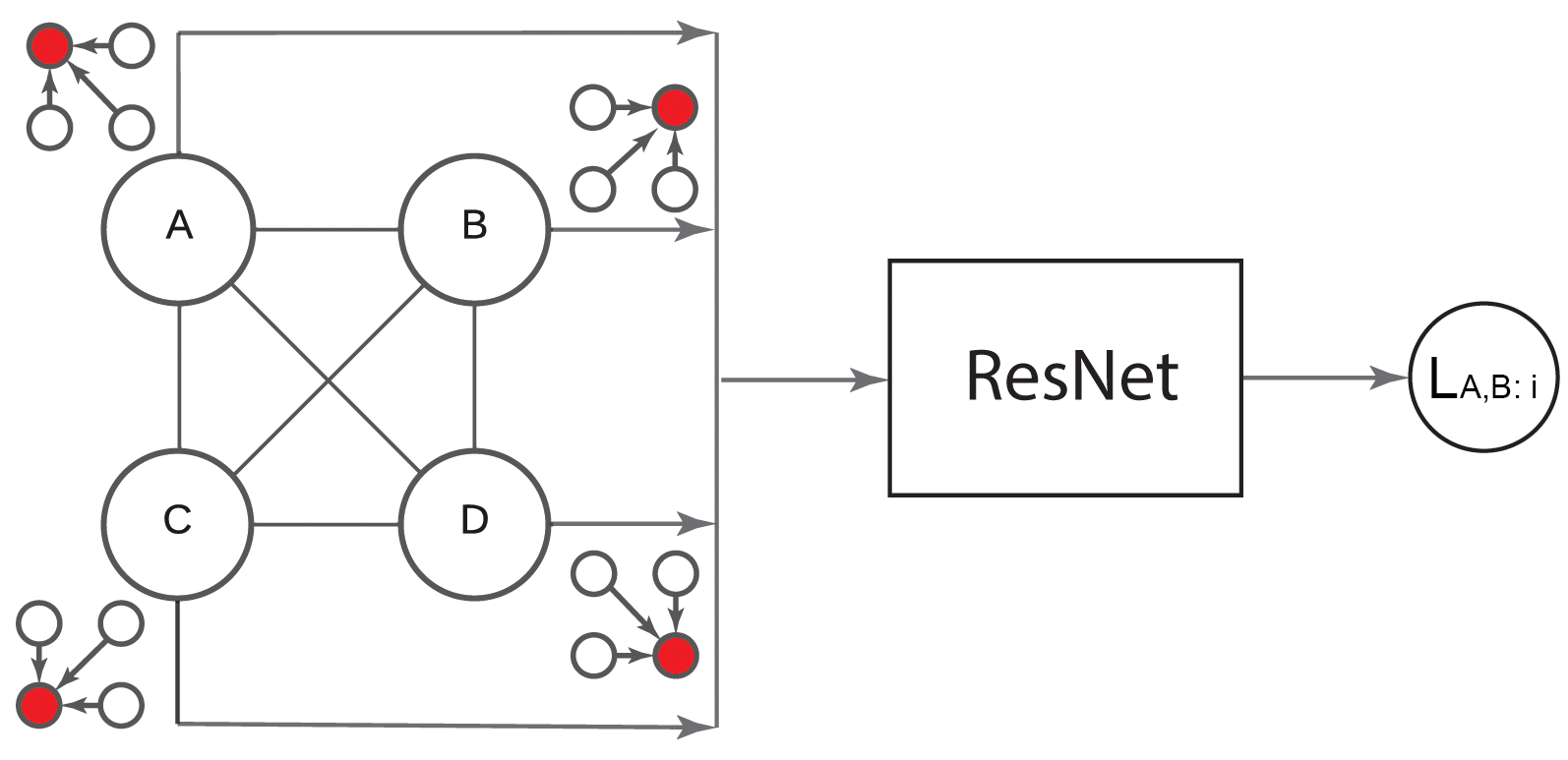} 
    \caption{GNN architecture for learning the $i$th-order temperature-independent excess parameter of binary liquids, denoted as $L_{A,B;i}$.}
    \label{Figures:architecture}
\end{figure} 

    In our neural network architecture, as shown in Figure \ref{Figures:architecture}, when a material (composed of its elements and composition) is inputted, each element is transformed into 14 distinct features, including atomic radius, atomic mass, electronegativity, core and valence electrons, ionization energy, electron affinity, density, and position in the periodic table. These features are then encoded and passed to the subsequent layer, which aims to capture the key determinants of the individual atomic contributions to the target variable. Furthermore, the elemental features interact with each other through the GNN connections, enabling contributions from binary, ternary, and higher-order element combinations. These encoded contributions are also propagated to the following layer. This layer, which encompasses unary, binary, and ternary interactions of the elements and compositions of the material, is then fed into a ResNet architecture, facilitating regression and estimation of the excess parameter.

\begin{figure}[]
    \centering
    \includegraphics[width=0.49\textwidth]{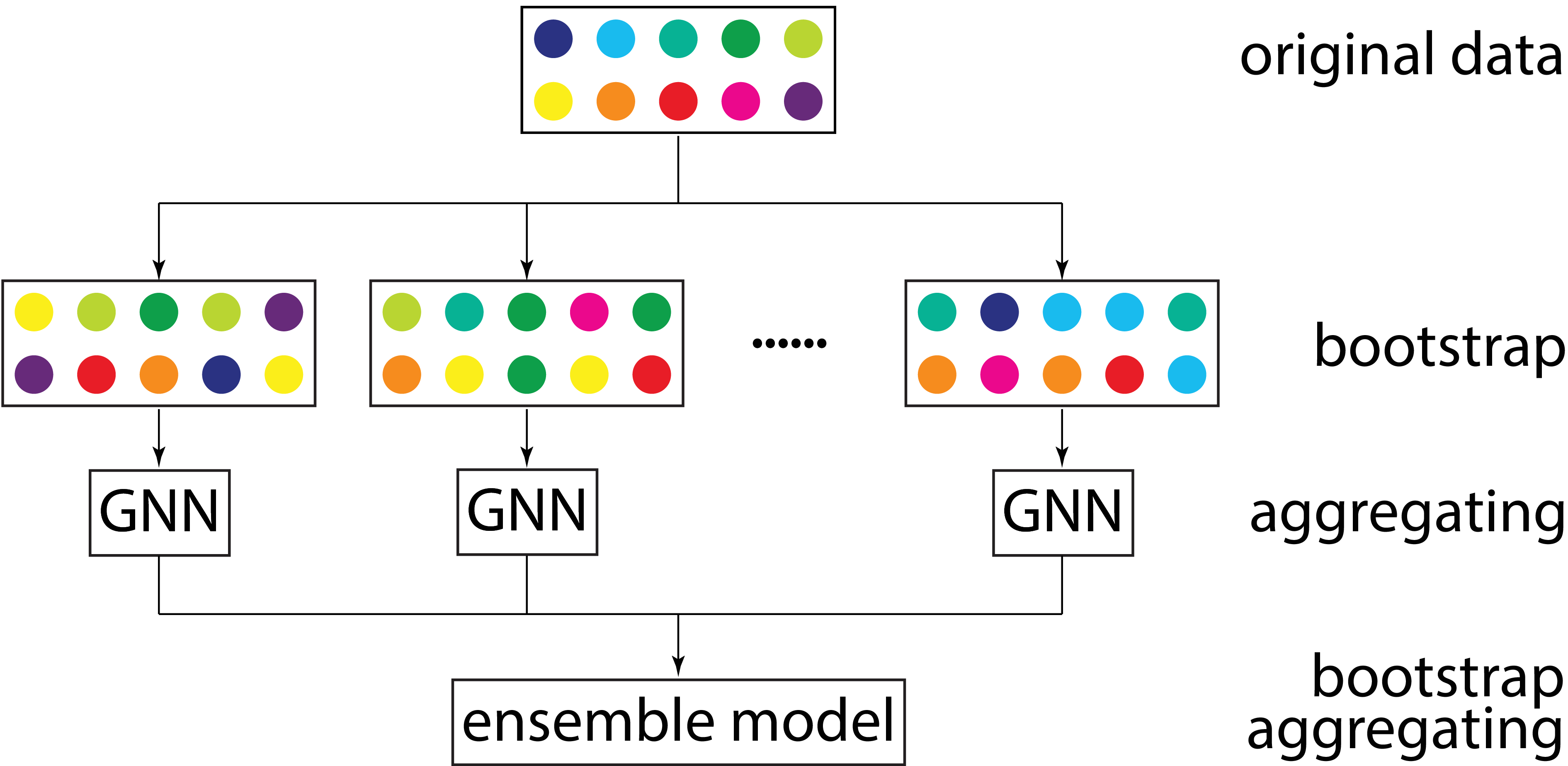} 
    \caption{The ensemble model of 30 GNN models is based on bootstrap aggregating. The benefits include (1) further overfitting reduction, (2) uncertainty estimate, and (3) outlier detection.}
    \label{Figures:ensemble}
\end{figure} 
We employ a system that incorporates database reshuffling to mitigate overfitting and estimate uncertainty. To achieve this, we employ bootstrap aggregating (bagging) \cite{Breiman1996} and construct an ensemble model comprising 30 GNN models, as depicted in Figure \ref{Figures:ensemble}. In comparison to the original GNN model, the ensemble model offers several advantages. It enables the estimation of computation uncertainty. Each chemical formula is processed independently by the 30 distinct GNN models, resulting in 30 different target values. From these values, we calculate the mean and standard error. Furthermore, the ensemble model serves as an additional safeguard against overfitting. The bootstrap technique reshuffles the dataset, generating randomized data subsets for training. Subsequently, the individual GNN models are aggregated to form the ensemble model, surpassing the performance of a single learner.

\begin{figure}[]
    \centering
    \includegraphics[width=0.49\textwidth]{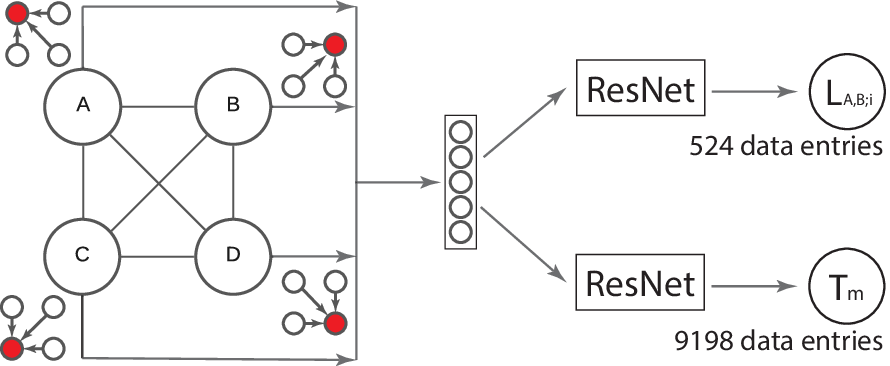} 
    \caption{Multi-task learning to improve the performance of the excess parameter model, by leveraging the large melting temperature dataset.}
    \label{Figures:multitask}
\end{figure} 

To enhance the performance of our model, we employ multi-task learning \cite{Caruana1997}. Given the limited availability of data, achieving high performance solely from one dataset is challenging. Therefore, we leverage a melting temperature dataset with a larger size and higher data quality to aid in improving the performance of a smaller dataset through multi-task learning. As illustrated in Figure \ref{Figures:multitask}, the GNN section of the two tasks is enforced to be identical, facilitating knowledge transfer from the melting temperature task to the CALPHAD parameter task. Meanwhile, the remaining ResNet section is allowed to be different, enabling independent learning of the two responses. This multi-task learning approach enhances our model's capabilities and overall performance.

\section{Results and Discussion \label{results}}

\begin{figure*}[]
    \centering
    \includegraphics[width=0.89\textwidth]{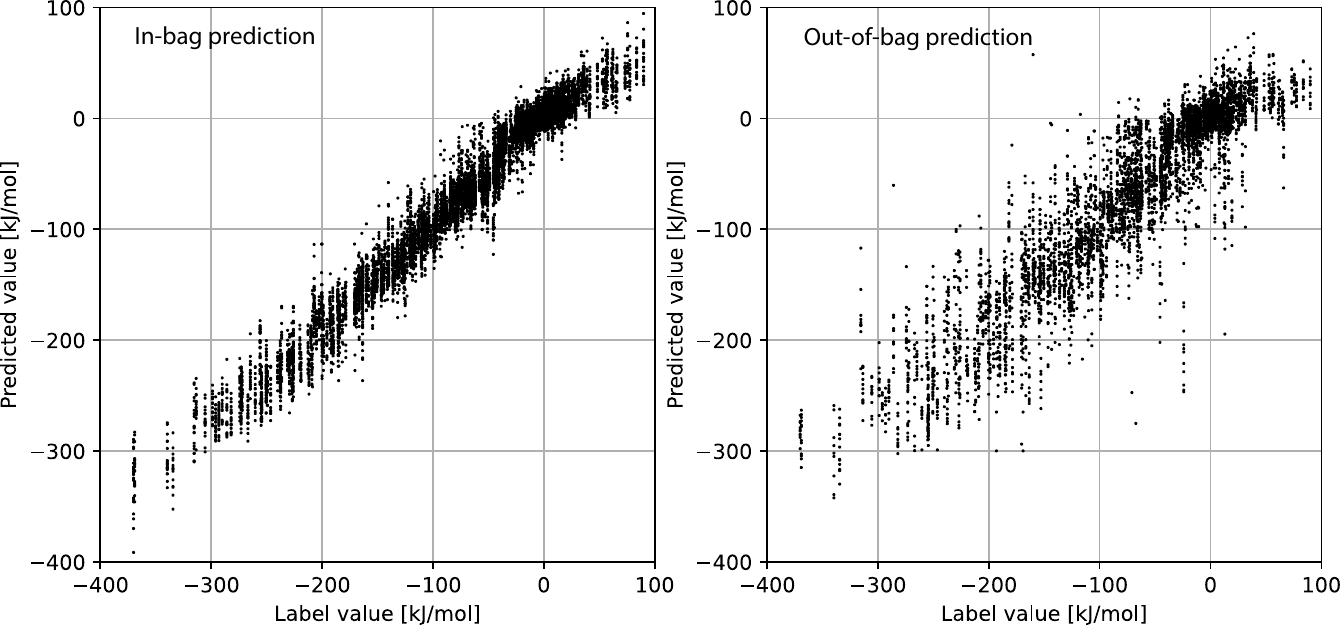} 
    \caption{Performance of multi-task learning for in-bag (training) and out-of-bag (testing) predictions. To form an ensemble model, we employ a process called bootstrapping where we sample 30 different subsets from the dataset. RMSE is 18 kJ/mol for in-bag (training) predictions and 35 kJ/mol for out-of-bag (testing) predictions, with R$^2$ score of 0.96 and 0.85, respectively.}
    \label{Figures:pred}
\end{figure*} 

Here we showcase the application of the GNN approach to construct a model for predicting the first-order temperature-independent excess parameter of binary liquids, denoted as $L_{A,B;0}$. The dataset used in our analysis was obtained from the TDBDB database \cite{vandewalle2018}, containing a total of 782 entries. After eliminating duplicates, the dataset was reduced to 581 entries. To streamline the analysis and avoid potential challenges associated with nonmetal elements (B, C, N, P, H), which necessitate a substantial amount of additional data, we excluded them from the dataset. As a result, the final dataset consisted of 524 data entries for analysis and model development.

Our ensemble model, consisting of 30 GNN models, achieves an RMSE (Root Mean Square Error) of 16 kJ/mol for in-bag (training) predictions and 38 kJ/mol for out-of-bag (testing) predictions, with an R$^2$ score of 0.97 and 0.82 respectively. These performance metrics indicate the model's accuracy and ability to generalize beyond the training data. This suggests that our model has the capability to predict the CALPHAD parameter, even for systems that are currently unknown to us.

To enhance the performance of our model, we employ the power of multi-task learning by training a model using both the CALPHAD parameter and melting temperature datasets. This multi-task approach yields significant improvements, with an RMSE of 18 kJ/mol for in-bag (training) predictions and 35 kJ/mol for out-of-bag (testing) predictions. The model achieves an R$^2$ score of 0.96 for in-bag predictions and 0.85 for out-of-bag predictions, demonstrating its enhanced performance and accuracy. These results highlight the effectiveness of the multi-task learning approach in improving the overall quality of our model.

The achievement of an R$^2$ score of 0.85 strongly validates our approach of utilizing a deep learning architecture to compute CALPHAD parameters solely based on the chemical formula. This success provides great encouragement for our future plans to expand our modeling efforts beyond the parameter $L_{A,B;0}$. We are excited to explore and develop additional models that leverage the power of deep learning for a broader range of materials-related predictions and analyses.

Although an R$^2$ score of 0.85 may not be considered high or ideal performance, it is important to note several reasons for this outcome. Firstly, the limited size of the CALPHAD parameter dataset presents challenges in training a robust model. To address this limitation, we plan to augment the dataset with additional data points from other databases, thereby enhancing the model's performance. Additionally, we intend to utilize SLUSCHI \cite{Hong2016}, our DFT package, to generate liquid mixing energy data, which will contribute to further improvements.

Secondly, it is crucial to acknowledge that CALPHAD parameters inherently possess certain limitations in terms of reliability. CALPHAD modelers construct models that best match the phase boundaries in their training data, ensuring their reliability. However, the parameters themselves may have inherent uncertainties, as they can exhibit similar bias across competing phases. Resolving this issue requires dedicated research that extends beyond the scope of this paper.

We will be deploying this model online on our group's website \cite{ML_model_webpage} as part of the Materials Properties Prediction (MAPP) framework. The model will be accessible to the public, allowing users to conveniently calculate the parameter by inputting the chemical formula. We believe that providing this accessible and user-friendly platform will greatly benefit the scientific community and enhance the ease of parameter calculations for various materials.

\section{Conclusion \label{conclusion}}
We apply our recently developed graph neural network approach to investigate CALPHAD parameters. Specifically, we demonstrate the successful learning of the first-order excess parameter of liquids ($L_{A,B;0}$). The effective acquisition of this parameter highlights the potential of employing this approach to obtain general CALPHAD parameters, thereby ultimately automating the construction of CALPHAD models through deep learning. Although beyond the scope of this work, we plan to further investigate this topic in our upcoming paper.

\bibliography{./bibliography.bib}

\begin{thebibliography}{16}%
\makeatletter
\providecommand \@ifxundefined [1]{%
 \@ifx{#1\undefined}
}%
\providecommand \@ifnum [1]{%
 \ifnum #1\expandafter \@firstoftwo
 \else \expandafter \@secondoftwo
 \fi
}%
\providecommand \@ifx [1]{%
 \ifx #1\expandafter \@firstoftwo
 \else \expandafter \@secondoftwo
 \fi
}%
\providecommand \natexlab [1]{#1}%
\providecommand \enquote  [1]{``#1''}%
\providecommand \bibnamefont  [1]{#1}%
\providecommand \bibfnamefont [1]{#1}%
\providecommand \citenamefont [1]{#1}%
\providecommand \href@noop [0]{\@secondoftwo}%
\providecommand \href [0]{\begingroup \@sanitize@url \@href}%
\providecommand \@href[1]{\@@startlink{#1}\@@href}%
\providecommand \@@href[1]{\endgroup#1\@@endlink}%
\providecommand \@sanitize@url [0]{\catcode `\\12\catcode `\$12\catcode
  `\&12\catcode `\#12\catcode `\^12\catcode `\_12\catcode `\%12\relax}%
\providecommand \@@startlink[1]{}%
\providecommand \@@endlink[0]{}%
\providecommand \url  [0]{\begingroup\@sanitize@url \@url }%
\providecommand \@url [1]{\endgroup\@href {#1}{\urlprefix }}%
\providecommand \urlprefix  [0]{URL }%
\providecommand \Eprint [0]{\href }%
\providecommand \doibase [0]{http://dx.doi.org/}%
\providecommand \selectlanguage [0]{\@gobble}%
\providecommand \bibinfo  [0]{\@secondoftwo}%
\providecommand \bibfield  [0]{\@secondoftwo}%
\providecommand \translation [1]{[#1]}%
\providecommand \BibitemOpen [0]{}%
\providecommand \bibitemStop [0]{}%
\providecommand \bibitemNoStop [0]{.\EOS\space}%
\providecommand \EOS [0]{\spacefactor3000\relax}%
\providecommand \BibitemShut  [1]{\csname bibitem#1\endcsname}%
\let\auto@bib@innerbib\@empty
%</preamble>
\bibitem [{\citenamefont {Kaufman}\ and\ \citenamefont
  {Bernstein}(1970)}]{Kaufman1970}%
  \BibitemOpen
  \bibfield  {author} {\bibinfo {author} {\bibfnamefont {L.}~\bibnamefont
  {Kaufman}}\ and\ \bibinfo {author} {\bibfnamefont {H.}~\bibnamefont
  {Bernstein}},\ }\href
  {http://inis.iaea.org/search/search.aspx?orig_q=RN:02004171} {\emph {\bibinfo
  {title} {Computer calculation of phase diagrams With special reference to
  refractory metals}}}\ (\bibinfo  {publisher} {Academic Press Inc},\ \bibinfo
  {year} {1970})\BibitemShut {NoStop}%
\bibitem [{\citenamefont {Lukas}\ \emph {et~al.}(1977)\citenamefont {Lukas},
  \citenamefont {Henig},\ and\ \citenamefont {Zimmermann}}]{Lukas1977}%
  \BibitemOpen
  \bibfield  {author} {\bibinfo {author} {\bibfnamefont {H.~L.}\ \bibnamefont
  {Lukas}}, \bibinfo {author} {\bibfnamefont {E.}~\bibnamefont {Henig}}, \ and\
  \bibinfo {author} {\bibfnamefont {B.}~\bibnamefont {Zimmermann}},\ }\href
  {\doibase https://doi.org/10.1016/0364-5916(77)90002-5} {\bibfield  {journal}
  {\bibinfo  {journal} {Calphad}\ }\textbf {\bibinfo {volume} {1}},\ \bibinfo
  {pages} {225} (\bibinfo {year} {1977})}\BibitemShut {NoStop}%
\bibitem [{\citenamefont {Marquardt}(1963)}]{Marquardt1963}%
  \BibitemOpen
  \bibfield  {author} {\bibinfo {author} {\bibfnamefont {D.~W.}\ \bibnamefont
  {Marquardt}},\ }\href {http://www.jstor.org/stable/2098941} {\bibfield
  {journal} {\bibinfo  {journal} {Journal of the Society for Industrial and
  Applied Mathematics}\ }\textbf {\bibinfo {volume} {11}},\ \bibinfo {pages}
  {431} (\bibinfo {year} {1963})}\BibitemShut {NoStop}%
\bibitem [{\citenamefont {Königsberger}(1991)}]{Konigsberger1991}%
  \BibitemOpen
  \bibfield  {author} {\bibinfo {author} {\bibfnamefont {E.}~\bibnamefont
  {Königsberger}},\ }\href {\doibase
  https://doi.org/10.1016/0364-5916(91)90027-H} {\bibfield  {journal} {\bibinfo
   {journal} {Calphad}\ }\textbf {\bibinfo {volume} {15}},\ \bibinfo {pages}
  {69} (\bibinfo {year} {1991})}\BibitemShut {NoStop}%
\bibitem [{\citenamefont {Hong}(2021)}]{hong2021melting}%
  \BibitemOpen
  \bibfield  {author} {\bibinfo {author} {\bibfnamefont {Q.-J.}\ \bibnamefont
  {Hong}},\ }\href@noop {} {\enquote {\bibinfo {title} {A melting temperature
  database and a neural network model for melting temperature prediction},}\ }
  (\bibinfo {year} {2021}),\ \Eprint {http://arxiv.org/abs/2110.10748}
  {arXiv:2110.10748 [cond-mat.mtrl-sci]} \BibitemShut {NoStop}%
\bibitem [{\citenamefont {Hong}\ \emph
  {et~al.}(2022{\natexlab{a}})\citenamefont {Hong}, \citenamefont {Ushakov},
  \citenamefont {Walle}, \citenamefont {Navrotsky}, \citenamefont {Burton},\
  and\ \citenamefont {Khvan}}]{Hong2022_pnas}%
  \BibitemOpen
  \bibfield  {author} {\bibinfo {author} {\bibfnamefont {Q.-J.}\ \bibnamefont
  {Hong}}, \bibinfo {author} {\bibfnamefont {S.~V.}\ \bibnamefont {Ushakov}},
  \bibinfo {author} {\bibfnamefont {A.~V.~D.}\ \bibnamefont {Walle}}, \bibinfo
  {author} {\bibfnamefont {A.}~\bibnamefont {Navrotsky}}, \bibinfo {author}
  {\bibfnamefont {B.}~\bibnamefont {Burton}}, \ and\ \bibinfo {author}
  {\bibfnamefont {A.}~\bibnamefont {Khvan}},\ }\href {\doibase 10.1073/pnas}
  {\bibfield  {journal} {\bibinfo  {journal} {Proceedings of the National
  Academy of Sciences}\ }\textbf {\bibinfo {volume} {119}},\ \bibinfo {pages}
  {e2209630119} (\bibinfo {year} {2022}{\natexlab{a}})}\BibitemShut {NoStop}%
\bibitem [{\citenamefont {Hong}(2022)}]{Hong2022_cms}%
  \BibitemOpen
  \bibfield  {author} {\bibinfo {author} {\bibfnamefont {Q.~J.}\ \bibnamefont
  {Hong}},\ }\href {\doibase 10.1016/j.commatsci.2022.111684} {\bibfield
  {journal} {\bibinfo  {journal} {Computational Materials Science}\ }\textbf
  {\bibinfo {volume} {214}} (\bibinfo {year} {2022}),\
  10.1016/j.commatsci.2022.111684}\BibitemShut {NoStop}%
\bibitem [{\citenamefont {Hong}\ \emph
  {et~al.}(2022{\natexlab{b}})\citenamefont {Hong}, \citenamefont {van~de
  Walle}, \citenamefont {Ushakov},\ and\ \citenamefont
  {Navrotsky}}]{Hong2022_calphad}%
  \BibitemOpen
  \bibfield  {author} {\bibinfo {author} {\bibfnamefont {Q.~J.}\ \bibnamefont
  {Hong}}, \bibinfo {author} {\bibfnamefont {A.}~\bibnamefont {van~de Walle}},
  \bibinfo {author} {\bibfnamefont {S.~V.}\ \bibnamefont {Ushakov}}, \ and\
  \bibinfo {author} {\bibfnamefont {A.}~\bibnamefont {Navrotsky}},\ }\href
  {\doibase 10.1016/j.calphad.2022.102500} {\bibfield  {journal} {\bibinfo
  {journal} {Calphad: Computer Coupling of Phase Diagrams and Thermochemistry}\
  }\textbf {\bibinfo {volume} {79}} (\bibinfo {year} {2022}{\natexlab{b}}),\
  10.1016/j.calphad.2022.102500}\BibitemShut {NoStop}%
\bibitem [{\citenamefont {Scarselli}\ \emph {et~al.}(2009)\citenamefont
  {Scarselli}, \citenamefont {Gori}, \citenamefont {Tsoi}, \citenamefont
  {Hagenbuchner},\ and\ \citenamefont {Monfardini}}]{Scarselli2009}%
  \BibitemOpen
  \bibfield  {author} {\bibinfo {author} {\bibfnamefont {F.}~\bibnamefont
  {Scarselli}}, \bibinfo {author} {\bibfnamefont {M.}~\bibnamefont {Gori}},
  \bibinfo {author} {\bibfnamefont {A.~C.}\ \bibnamefont {Tsoi}}, \bibinfo
  {author} {\bibfnamefont {M.}~\bibnamefont {Hagenbuchner}}, \ and\ \bibinfo
  {author} {\bibfnamefont {G.}~\bibnamefont {Monfardini}},\ }\href {\doibase
  10.1109/TNN.2008.2005605} {\bibfield  {journal} {\bibinfo  {journal} {IEEE
  Transactions on Neural Networks}\ }\textbf {\bibinfo {volume} {20}},\
  \bibinfo {pages} {61} (\bibinfo {year} {2009})}\BibitemShut {NoStop}%
\bibitem [{\citenamefont {He}\ \emph {et~al.}(2016)\citenamefont {He},
  \citenamefont {Zhang}, \citenamefont {Ren},\ and\ \citenamefont
  {Sun}}]{He2016}%
  \BibitemOpen
  \bibfield  {author} {\bibinfo {author} {\bibfnamefont {K.}~\bibnamefont
  {He}}, \bibinfo {author} {\bibfnamefont {X.}~\bibnamefont {Zhang}}, \bibinfo
  {author} {\bibfnamefont {S.}~\bibnamefont {Ren}}, \ and\ \bibinfo {author}
  {\bibfnamefont {J.}~\bibnamefont {Sun}}\ }(\bibinfo {year} {2016})\ pp.\
  \bibinfo {pages} {770--778}\BibitemShut {NoStop}%
\bibitem [{\citenamefont {Abadi}\ \emph {et~al.}(2015)\citenamefont {Abadi},
  \citenamefont {Agarwal}, \citenamefont {Barham}, \citenamefont {Brevdo},
  \citenamefont {Chen}, \citenamefont {Citro}, \citenamefont {Corrado},
  \citenamefont {Davis}, \citenamefont {Dean}, \citenamefont {Devin},
  \citenamefont {Ghemawat}, \citenamefont {Goodfellow}, \citenamefont {Harp},
  \citenamefont {Irving}, \citenamefont {Isard}, \citenamefont {Jia},
  \citenamefont {Jozefowicz}, \citenamefont {Kaiser}, \citenamefont {Kudlur},
  \citenamefont {Levenberg}, \citenamefont {Mané}, \citenamefont {Monga},
  \citenamefont {Moore}, \citenamefont {Murray}, \citenamefont {Olah},
  \citenamefont {Schuster}, \citenamefont {Shlens}, \citenamefont {Steiner},
  \citenamefont {Sutskever}, \citenamefont {Talwar}, \citenamefont {Tucker},
  \citenamefont {Vanhoucke}, \citenamefont {Vasudevan}, \citenamefont
  {Viégas}, \citenamefont {Vinyals}, \citenamefont {Warden}, \citenamefont
  {Wattenberg}, \citenamefont {Wicke}, \citenamefont {Yu},\ and\ \citenamefont
  {Zheng}}]{tensorflow}%
  \BibitemOpen
  \bibfield  {author} {\bibinfo {author} {\bibfnamefont {M.}~\bibnamefont
  {Abadi}}, \bibinfo {author} {\bibfnamefont {A.}~\bibnamefont {Agarwal}},
  \bibinfo {author} {\bibfnamefont {P.}~\bibnamefont {Barham}}, \bibinfo
  {author} {\bibfnamefont {E.}~\bibnamefont {Brevdo}}, \bibinfo {author}
  {\bibfnamefont {Z.}~\bibnamefont {Chen}}, \bibinfo {author} {\bibfnamefont
  {C.}~\bibnamefont {Citro}}, \bibinfo {author} {\bibfnamefont {G.~S.}\
  \bibnamefont {Corrado}}, \bibinfo {author} {\bibfnamefont {A.}~\bibnamefont
  {Davis}}, \bibinfo {author} {\bibfnamefont {J.}~\bibnamefont {Dean}},
  \bibinfo {author} {\bibfnamefont {M.}~\bibnamefont {Devin}}, \bibinfo
  {author} {\bibfnamefont {S.}~\bibnamefont {Ghemawat}}, \bibinfo {author}
  {\bibfnamefont {I.}~\bibnamefont {Goodfellow}}, \bibinfo {author}
  {\bibfnamefont {A.}~\bibnamefont {Harp}}, \bibinfo {author} {\bibfnamefont
  {G.}~\bibnamefont {Irving}}, \bibinfo {author} {\bibfnamefont
  {M.}~\bibnamefont {Isard}}, \bibinfo {author} {\bibfnamefont
  {Y.}~\bibnamefont {Jia}}, \bibinfo {author} {\bibfnamefont {R.}~\bibnamefont
  {Jozefowicz}}, \bibinfo {author} {\bibfnamefont {L.}~\bibnamefont {Kaiser}},
  \bibinfo {author} {\bibfnamefont {M.}~\bibnamefont {Kudlur}}, \bibinfo
  {author} {\bibfnamefont {J.}~\bibnamefont {Levenberg}}, \bibinfo {author}
  {\bibfnamefont {D.}~\bibnamefont {Mané}}, \bibinfo {author} {\bibfnamefont
  {R.}~\bibnamefont {Monga}}, \bibinfo {author} {\bibfnamefont
  {S.}~\bibnamefont {Moore}}, \bibinfo {author} {\bibfnamefont
  {D.}~\bibnamefont {Murray}}, \bibinfo {author} {\bibfnamefont
  {C.}~\bibnamefont {Olah}}, \bibinfo {author} {\bibfnamefont {M.}~\bibnamefont
  {Schuster}}, \bibinfo {author} {\bibfnamefont {J.}~\bibnamefont {Shlens}},
  \bibinfo {author} {\bibfnamefont {B.}~\bibnamefont {Steiner}}, \bibinfo
  {author} {\bibfnamefont {I.}~\bibnamefont {Sutskever}}, \bibinfo {author}
  {\bibfnamefont {K.}~\bibnamefont {Talwar}}, \bibinfo {author} {\bibfnamefont
  {P.}~\bibnamefont {Tucker}}, \bibinfo {author} {\bibfnamefont
  {V.}~\bibnamefont {Vanhoucke}}, \bibinfo {author} {\bibfnamefont
  {V.}~\bibnamefont {Vasudevan}}, \bibinfo {author} {\bibfnamefont
  {F.}~\bibnamefont {Viégas}}, \bibinfo {author} {\bibfnamefont
  {O.}~\bibnamefont {Vinyals}}, \bibinfo {author} {\bibfnamefont
  {P.}~\bibnamefont {Warden}}, \bibinfo {author} {\bibfnamefont
  {M.}~\bibnamefont {Wattenberg}}, \bibinfo {author} {\bibfnamefont
  {M.}~\bibnamefont {Wicke}}, \bibinfo {author} {\bibfnamefont
  {Y.}~\bibnamefont {Yu}}, \ and\ \bibinfo {author} {\bibfnamefont
  {X.}~\bibnamefont {Zheng}},\ }\href {https://www.tensorflow.org/} {\enquote
  {\bibinfo {title} {Tensorflow: Large-scale machine learning on heterogeneous
  systems},}\ } (\bibinfo {year} {2015}),\ \bibinfo {note} {software available
  from tensorflow.org}\BibitemShut {NoStop}%
\bibitem [{\citenamefont {Breiman}(1996)}]{Breiman1996}%
  \BibitemOpen
  \bibfield  {author} {\bibinfo {author} {\bibfnamefont {L.}~\bibnamefont
  {Breiman}},\ }\href {\doibase 10.1007/BF00058655} {\bibfield  {journal}
  {\bibinfo  {journal} {Machine Learning}\ }\textbf {\bibinfo {volume} {24}},\
  \bibinfo {pages} {123} (\bibinfo {year} {1996})}\BibitemShut {NoStop}%
\bibitem [{\citenamefont {Caruana}(1997)}]{Caruana1997}%
  \BibitemOpen
  \bibfield  {author} {\bibinfo {author} {\bibfnamefont {R.}~\bibnamefont
  {Caruana}},\ }\href {\doibase 10.1023/A:1007379606734} {\bibfield  {journal}
  {\bibinfo  {journal} {Machine Learning}\ }\textbf {\bibinfo {volume} {28}},\
  \bibinfo {pages} {41} (\bibinfo {year} {1997})}\BibitemShut {NoStop}%
\bibitem [{\citenamefont {van~de Walle}\ \emph {et~al.}(2018)\citenamefont
  {van~de Walle}, \citenamefont {Nataraj},\ and\ \citenamefont
  {Liu}}]{vandewalle2018}%
  \BibitemOpen
  \bibfield  {author} {\bibinfo {author} {\bibfnamefont {A.}~\bibnamefont
  {van~de Walle}}, \bibinfo {author} {\bibfnamefont {C.}~\bibnamefont
  {Nataraj}}, \ and\ \bibinfo {author} {\bibfnamefont {Z.-K.}\ \bibnamefont
  {Liu}},\ }\href {\doibase https://doi.org/10.1016/j.calphad.2018.04.003}
  {\bibfield  {journal} {\bibinfo  {journal} {Calphad}\ }\textbf {\bibinfo
  {volume} {61}},\ \bibinfo {pages} {173} (\bibinfo {year} {2018})}\BibitemShut
  {NoStop}%
\bibitem [{\citenamefont {Hong}\ and\ \citenamefont {van~de
  Walle}(2016)}]{Hong2016}%
  \BibitemOpen
  \bibfield  {author} {\bibinfo {author} {\bibfnamefont {Q.-J.}\ \bibnamefont
  {Hong}}\ and\ \bibinfo {author} {\bibfnamefont {A.}~\bibnamefont {van~de
  Walle}},\ }\href {\doibase 10.1016/j.calphad.2015.12.003} {\bibfield
  {journal} {\bibinfo  {journal} {Calphad: Computer Coupling of Phase Diagrams
  and Thermochemistry}\ }\textbf {\bibinfo {volume} {52}},\ \bibinfo {pages}
  {88} (\bibinfo {year} {2016})}\BibitemShut {NoStop}%
\bibitem [{ML_()}]{ML_model_webpage}%
  \BibitemOpen
  \href
  {https://faculty.engineering.asu.edu/hong/melting-temperature-predictor/}
  {\enquote {\bibinfo {title} {Melting temperature predictor based on machine
  learning},}\ }\BibitemShut {NoStop}%
\end{thebibliography}%
%\printbibliography %Prints bibliography

\end{document}